\documentclass[superscriptaddress,preprint,secnumarabic,amssymb,amsmath,nobibnotes,aps,prd,showpacs,nofootinbib,showkeys]{revtex4}

\usepackage{graphicx}
\usepackage{epsf}
\usepackage{bm}
\usepackage{amsmath}
\usepackage{amsfonts}
\usepackage{amssymb}
\usepackage{epstopdf}
\usepackage{natbib}
\usepackage{color}%
\providecommand{\U}[1]{\protect\rule{.1in}{.1in}}
\newcommand{\bee}{\begin{equation}}
\newcommand{\eee}{\end{equation}}
\newcommand{\baa}{\begin{eqnarray}}
\newcommand{\eaa}{\end{eqnarray}}

\newcommand{\mincir}{\raise
-3.truept\hbox{\rlap{\hbox{$\sim$}}\raise4.truept\hbox{$<$}\ }}
\newcommand{\magcir}{\raise
-3.truept\hbox{\rlap{\hbox{$\sim$}}\raise4.truept\hbox{$>$}\ }}

\usepackage{color}
\def\ni{\noindent}

\begin{document}

\title{\Large The magnetocaloric effect from the point of view \\ of Tsallis non-extensive thermostatistics}

\author{Isa\' ias G. de Oliveira}
\email{isaias@ufrrj.br}
\affiliation{Grupo de F\'isica Te\'orica e Matem\'atica F\'isica, Departamento de F\'isica,
Universidade Federal Rural do Rio de Janeiro, 23897-000 Serop\'edica, Rio de Janeiro, Brazil}

\author{Mois\'es A. S. M. Ara\'ujo}
\email{mmaraujo@ufrrj.br}
\affiliation{Grupo de F\'isica Te\'orica e Matem\'atica F\'isica, Departamento de F\'isica,
Universidade Federal Rural do Rio de Janeiro, 23897-000 Serop\'edica, Rio de Janeiro, Brazil}

\author{Everton M. C. Abreu}
\email{evertonabreu@ufrrj.br}
\affiliation{Grupo de F\'isica Te\'orica e Matem\'atica F\'isica, Departamento de F\'isica,
Universidade Federal Rural do Rio de Janeiro, 23897-000 Serop\'edica, Rio de Janeiro, Brazil}
\affiliation{Departamento de F\'isica, Universidade Federal de Juiz de Fora, 36036-330, Juiz de Fora, MG, Brazil}

\author{Mario J. Neves}
\email{mariojr@ufrrj.br}
\affiliation{Grupo de F\'isica Te\'orica e Matem\'atica F\'isica, Departamento de F\'isica,
Universidade Federal Rural do Rio de Janeiro, 23897-000 Serop\'edica, Rio de Janeiro, Brazil}

\date{\today}


%%%%%%%%%%%%%%
\begin{abstract}
\ni   In this work we have analyzed the magnetocaloric effect (MCE) from the Tsallis thermostatistics formalism (TTF) point of view.  The problem discussed here is a two level system MCE.   We have calculated, both analytically and numerically, the entropy of this system as a function of the  Tsallis' parameter (the well known $q$-parameter) which value depends on the extensivity ($q<1$) or non-extensivity ($q>1$) of the system. Since we consider this MCE not depending on the initial conditions, which classify our system as a non-extensive one, we used   
several greater than one $q$-parameters to understand the effect of the nonextensive formalism in the entropy as well as the magnetocaloric potential, $\Delta S$.  We have plotted several curves that shows precisely the behavior of this effect when dealt with non-extensive statistics.
\end{abstract}

%%%%%%%%%%%%%%%%%%%%%%%%%%%%%%%%%%%%%%%%%%%%%%%%%%%%%%%%%%%%%%%%%%
\maketitle
%%%%%%%%%%%%%%%%%%%%%%%%%%%%%%%%%%%%%%%%%%%%%%%%%%%%%%%%%%%%%%%%%%

\section{Introduction}

Approximately thirty years ago, Tsallis formulated a  generalization of the Boltzmann-Gibbs (BG) statistics for non-extensive systems and the results were extremely satisfactory.  As a result of this success, the formalism has begun to be used in several and completely different areas of research  \cite{tsallis,tsallis2,tsallis3}, which will be exemplified in the near future.  

Non-extensivity is a property of the systems where long-range
interactions, space-time complexity or independence of the initial conditions are present.
Long-range forces can be found in astrophysical and
nano (length) scales. Space-time complexity means that we can
depict the presence of long-range space
and time correlations.  It is detected in equilibrium statistical
mechanics to emerge at critical points for second order
phase transitions. Moreover, the concept of self-organized
criticality was recently introduced to describe driven
systems which naturally develop to a dynamical attractor
ready at criticality \cite{btw}. Self-organized criticality is
defined as being in the origin of fractal constructions, anomalous diffusion, noise
with a $1/f$ power spectrum,  L\'evy
flights, and punctuated equilibrium behavior \cite{pmb}, which are
features of the non-extensive character of the dynamics
attractor \cite{lt}.

Also known as Tsallis thermostatistics formalism (TTF), several areas of research have received relevant new numerical and analytical analysis concerning some of its main problems, which could not find solutions using the standard BG approach, where we have $q=1$. As examples, we can mention that such new non-extensive calculations have been performed in many completely different subjects, such as earthquakes
\cite{3}, chaos \cite{4}, tracers on random systems \cite{5},
random quenched and fractals \cite{6,8}, specific heat \cite{7},
clusters \cite{9}, growth models \cite{10}, econophysics issues such as stock markets \cite{11} and income distribution \cite{meu,marcelo},   the Levy-type anomalous diffusion \cite{levy}, turbulence in a pure-electron plasma \cite{turb} and gravitational systems \cite{sys},  inverse {\it bremsstrahlung} in plasma \cite{ts}
and non-extensive statistical mechanics log-periodic
oscillations \cite{12}, to mention only some of them.

\bigskip

The experimental discovery of the giant magnetocaloric effect (MCE) by Pecharsky and Gschneidner \cite{pg}, has motivated several research groups to investigate the microscopic mechanisms that rule such effect.  The magnetocaloric effect is the capacity that the material has to lower its temperature when it is submitted to an external increasing magnetic field, and to raise its temperature when the applied magnetic field diminishes.

The main motivation to carry out today investigations concerning the MCE resides in its utilization as an alternative technology consideration for  refrigeration from an ambient temperature down to the hydrogen and helium liquefaction temperature.   The effect replaces the usual gas expansion - compression today's technology.  It is well known that materials with the larger MCEs are necessary in order to enhance the energy efficiency.

It is also well known that the majority of MCE analysis were accomplished on ferromagnetic materials near their respective Curie temperature, namely, ferromagnetic on paramagnetic transitions.   The objective here is to introduce a new parameter, the $q$-parameter, to help to establish the magnetocaloric properties of such materials.  We believe that the different values of $q$ explain different behavior concerning MCE materials.   Of course we do not expect to explain all the features shown by these materials.   Our effort here is to explain the possibility of the existence of a new parameter in the MCE scenario, the $q$-parameter, by analyzing its behavior as depending on this parameter.  Several $q$'s were used here under different circumstances to show this behavior.

Theoretically speaking, when a magnetic material is magnetized, the spins try to line up with the field.   This line-up effect causes a decreasing of the entropy.   When the magnetic field is removed, the spins are guided to a random order, increasing in this way the entropy of the material.   The behavior of this difference concerning the entropy is called $\Delta S$, which is a function of the temperature and a function of the applied field. It will be analyzed here for different values of $q$.   

Statistically speaking, we will analyze a microscopic system that has two levels of energy, separated by an energy gap given by $\delta$.   When $k_B\,T << \delta$, the higher level of energy turns out to be almost empty.   On the other hand, for $k_B\,T >> \delta$, both levels are equally occupied.   The transition from the lower level to the excited one occurs with higher probability for thermal energies which have energy around $\delta$.   This sudden internal energy variation corresponds to a peak in the specific heat and goes to zero for both higher and lower temperatures.   This effect is known as the Schottky effect for the specific heat.

In this work we have obeyed the following organization: in section 2 we described the main ingredients of the standard Tsallis formalism.     The results and discussions were given in section 3.   The conclusions are given in section 4.  
%To turn the comparison between the curves obtained here easier, we have chosen to show all the graphs together at the end of the paper, after the references list.

%%%%%%%%%%%%%%%%%%%%%%%%%%%%%%%%%%%%%%%%%%%%%%%%%%%%%%%%%%%%%%%%%%%%%%%%%%%%%%%%%%%%%%%%%%%%%%%%%%%%%%%%%%%%%%%%%%%%%%%%%%%%%%%%%%%%%%%%%%%%%%%%%%%%%%%%%%%%%%%%%%%%%%%%%%%%%%%%%%%%%%%%%%%%%%%%%%%%%%%%%%%%%%%%%%%%%%%%%%%%%%%%%%%%%%%%%%%%%%%%%%%%%%%%%%%%%%%%%%%%%%%%%%%%%%%%%%%%%%%%%%%%%%%%%%%%%%%%%%%%%%

\section{Tsallis thermostatistics formalism}
\label{ttf}
In a nutshell, Tsallis' thermostatistics \cite{tsallis}, which is a generalization of the Boltzman-Gibbs's (BG) statistical theory, defines a nonadditive entropy as
\begin{eqnarray}
\label{nes}
S_q =  k_B \, \frac{1 - \sum_{i=1}^W p_i^q}{q-1}\;\;\;\;\;\;\qquad\qquad \Big(\,\sum_{i=1}^W p_i = 1\,\Big)\,\,,
\end{eqnarray}

\ni where $p_i$ is the probability of a system to exist within a microstate, $W$ is the total number of configurations and
$q$ ($q \in \mathbb{R}$), known in the current literature as being the Tsallis parameter or the non-extensive  parameter, is a real parameter\footnote{Although there are some considerations about a complex $q$-parameter in the non-extensive literature \cite{meu,poloneses}} which measures the degree of non-extensivity according to the entropy relation
\bee
\label{2}
S_q (A+B)\,=\,S_q (A) + S_q (B)\,+\, (1-q)\,S_q (A) S_q(B)\,\,,
\eee

\ni where $A$ and $B$ are two independent systems.  Superextensivity (more probable events) corresponds to a $q > 1$, and subextensivity (rare events) corresponds to a $q < 1$.  Negative values for $q$ reflects a negative entropy, an ``order" possibility for the system, since the positive entropy is connected to the disorder tendency of a system.
The concept of extensivity is relative to the limit $q \rightarrow 1$.   The $q=1$ case corresponds to the extensive one, i.e., the BG statistics. It can be seen from Eq. \eqref{2}, that the standard additivity of entropy can be recovered.   The entropy is given by the well known $S\,=\, k_B \sum_{i=1} p_i \log\,p_i$.
The definition of entropy in TTF carries the standard properties of positivity, equiprobability, concavity and irreversibility.
%It is noteworthy to mention that the TTF has the BG statistics as a particular case in the limit $ q \rightarrow 1$ where the standard additivity of entropy can be recovered.

In the microcanonical ensemble, where all the states have the same probability, Tsallis entropy reduces to \cite{te}
\begin{eqnarray}
\label{micro}
S_q=k_B\, \frac{W^{1-q}-1}{1-q}\,\,,
\end{eqnarray}

\ni where in the limit $q \rightarrow 1$ we also recover the usual Boltzmann entropy formula, $S=k_B\, \ln {W}$.

In the context of Tsallis's general theory, the probability distribution can be modified in which the Boltzmann $e^{-\beta \, E_{n}}$ can be written as
\begin{eqnarray}
\Big[ \, 1+ (1-q) \, \beta \, E_{n} \, \Big]^{\frac{1}{q-1}} \,\, ,
\end{eqnarray}
where $\beta=(k_{B} \, T)^{-1}$. Thereby, the probability of a state $n$ with energy $E_n$ is given by
\begin{eqnarray}
p_{n}=Z_{q}^{-1}\Big[ \, 1+ (q-1) \, \beta \, E_{n} \, \Big]^{\frac{1}{q-1}} \; ,
\end{eqnarray}
where the partition function is
\begin{eqnarray}
Z_{q}=\sum_{n} \Big[ \, 1- (q-1) \, \beta \, E_{n} \, \Big]^{\frac{1}{q-1}} \; .
\end{eqnarray}
The internal energy is $U=\langle E_{n} \rangle$, where $\langle \, \cdots \, \rangle$ is defined by the average under the distribution.
For the two levels system with $N$-particles, the internal energy is given by
\begin{eqnarray}
U=N \, \Big(p_{0} \, E_{0} + p_{1} \, E_{1}\Big) \; .
\end{eqnarray}
We will fix the value $E_{0}\,=\,0$ for the ground state, and the energy of the excited state $E_{1}=E$, the internal energy can be written as
\begin{eqnarray}
\label{UU}
\frac{U}{N}=E \, \frac{[ \, 1+ (1-q) \, \beta \, E \, ]^{\frac{1}{q-1}}}{1+[ \, 1+ (1-q) \, \beta \, E \, ]^{\frac{1}{q-1}}} \; .
\end{eqnarray}
The specific heat is given by
\begin{eqnarray}
\label{CC}
C=\frac{\partial U}{\partial T}=- \, k_{B} \, \beta^{2} \, \frac{\partial U}{\partial \beta} \; ,
\end{eqnarray}
{\it i. e.},
\begin{eqnarray}
\label{CC2}
\frac{C}{N}=- \, k_{B} \, \beta^{2} \,\Big(1- \frac UN \Big) \, \frac{U-E}{1-(1-q) \, \beta \, E} \; .
\end{eqnarray}
Concerning Eq. \eqref{CC2}, when $q=1$ the BG statistic result can be obtained
for any $k\, T$-value. The specific heat properties can be discussed in the context of the Tsallis theory in \cite{ito,turb}.
%

%
%%%%%%%%%%%%%%%%%%%%%%%%%%%%%%%%%%%%%%%%%%%%%%%%%%%%%%%%%%%%%%%%%%%%%%%%%%%%%%%%%%%%%%%%%%%%%%%%%%%%%%%%%%%%%%%%%%%%%%%%%%%%%%%%%%%%%%%%%%%%%%%%%%%%%%%%%%%%%%%%%%%%%%%%%%%%%%%%%%%%%%%%%%%%%%%%%%%%%%%%%%%%%%%%%%%%%%%%%%%%%%%%%%%%%%%%%%%%%%%%%%%%%%%%%%%%%%%%%%%%%%%%%%%%%%%%%%%%%%%%%%%%%%%%%%%%%%%%%%%%%%
%
\section{Results and Discussion}

In this section we present the results that arise from the numerical solution of the equations introduced in the section above. Two quantities will be used to characterize the magnetocaloric effect. One of them is the magnetocaloric potential, $\Delta S$, defined below, and the another one is the isentropic temperature change $\Delta T_{ad}$. We will calculate the magnetocaloric potential $\Delta S$ for several values of Tsallis parameter $q$. The quantity $\Delta S$, introduced here is related to the isothermal process given by
\begin{eqnarray}
\label{SS}
-\Delta S=S(H=0,T,q)-S(H\neq0,T,q) \; .
\end{eqnarray}

In both graphics of Figure 1 we can see clearly the behavior of the specific heat  as a function of both the temperature and $q$ for different energy intervals.  Notice that for all $q$-values in the interval $1 < q < 2$ we have the same asymptotic behavior associated with the temperature.   We consider that this behavior is a fingerprint of the specific heat of a two-level quantum system.   It is the well known Schottky effect of the specific heat. In this range for the parameter $q$, as the extensivity of the system increase, the maximum value of the specific heat and the temperature of this maximum also increase. The fact that we have no numerical results for low temperatures is a consequence of a kind of forbidden region in Tsallis generalized formalism, for more details see \cite{tsallis,tsallis2,tsallis3}. It means that, as the $q$-values increases, the information about the entropy is lost for low temperatures, as can be seen directly in the graphic.

The curves in both Figures 2 and 3 show the entropy of the two level quantum system when a magnetic field is applied and  the other curve shows the scenario where there is no field at all. We can note that the behavior of these entropies, for values of $q$ near to one, presents similar behavior as the entropy calculated by BG statistic, which is a different behavior for values of $q$ far from one.   For larger values of $q$, and for low values of the temperature, it begins to appear a determined loss of information concerning the entropy, which is an intrinsic characteristic of the TTF, mentioned above.  The value of the area inside both curves represents the heat involved to take the system from the thermodynamic state $(H=0,T=0)$ to the state $(H\neq0,T)$. 
%We calculated both curves for some values of the $q$-parameters. 

Figure 4 shows the MC potential $\Delta S$ as function of the temperature, calculated through Eq. (\ref{SS}). The applied magnetic field used for all these curves is the magnetic field which split the two level system through a gap of energy $\delta = 1$ in arbitrary units. It is clear that the increasing of the extensivity of the system, i.e., with the increasing of $q$, we have the increasing of the maximum value of $\Delta S$, and there is a small shift in the temperature related to this maximum. One of the main characteristic of a good refrigerator material is to produce high values of $\Delta S$. Therefore, our results show that the extensivity of the system is an important parameter to take into account, in order to find out a good refrigerator material. Another important characteristic of the refrigerator material concerns the temperature of the maximum value of $\Delta S$, and as one can see, the extensivity of the system increase its temperature. In the present case, we can observe that a small increasing in the $q$-parameter generates an increasing in the MCE potential. For instance, from $q=1.01$ to $q=1.22$, the maximum of the MCE potential varies from $\Delta S = 0.05$ to $\Delta S = 0.12$, where the relative changes are $\Delta q/q=0.21$ and $\Delta S/S=1.4$, it is showing that the increasing of the $\Delta S$ is $6.7$ larger than the increasing of $\Delta q$.
It is important to note that although the heat $\Delta Q$, involved in the MCE is approximated equal for the range of the values of $q$ studied here, the values of the magnetocaloric potential $\Delta S$ have a large increasing.

To connect $q$ to a measured quantity we have constructed the graphic presented in Figure 5. In this figure one can see that the heat involved in the same process for different values of the $q$-parameter has approximately equal values.   This result shows that the extensivity of the system is not related to the heat of the process.

Finally, in Figure 6 we have shown the behavior of the internal energy according the temperature in arbitrary units for varying $q$-parameters and for different energies, $E=5.0$ and $E=6.0$, respectively.
%
%\begin{table}[ht]
%%\caption{Relation between $q$ and the area below the entropy curves} % title of Table
%\centering % used for centering table
%\begin{tabular}{ |p{2.5cm}|p{2.5cm}|p{2.5cm}|p{2.5cm}|  }
%%{c c c} % centered columns (3 columns)
%\hline\hline %inserts double horizontal lines
%$\:\quad\:\quad q$ & $\:\quad\:\quad T^*$ & $\:\quad\:\quad Area$ \\ [0.5ex] % inserts table
%%heading
%\hline\hline % inserts single horizontal line
%\:\quad\:\quad 1.01 & \:\quad\:\quad 2.17947 & \:\quad\:\quad 0.43192 \\ % inserting body of the table
%\:\quad\:\quad 1.1 & \:\quad\:\quad 2.33118 & \:\quad\:\quad 0.43761  \\
%\:\quad\:\quad 1.3 & \:\quad\:\quad 2.84067 & \:\quad\:\quad 0.42122  \\
%\:\quad\:\quad 1.6 & \:\quad\:\quad 3.67891 & \:\quad\:\quad 0.41236  \\[1ex]
%%\:\quad\:\quad 5 & \:\quad\:\quad 45 & \:\quad\:\quad 30  \\ [1ex] % [1ex] adds vertical space
%\hline %inserts single line
%\end{tabular}
%\caption{Relation between $q$ and the area below the entropy curves}
%\label{table:nonlin} % is used to refer this table in the text
%\end{table}

\section{Conclusion}

We have studied the magnetocaloric effect of the two level quantum system using the Tsallis thermodynamic statistic. We have calculated the magnetocaloric potential $\Delta S$ as a function of the temperature for equal values of the $q$-parameter. Our result shows that, although the maximum value of $\Delta S$ is very sensible to the value of $q$, the heat involved in the MCE is not. Our results suggest a substantial impact of the non-extensivity on the system in order to map an optimum magnetocaloric material. 
%To quantify this impact, we have construct several graphics where the behavior of all these parameters can be analyzed.

%In conclusion, we have shown how the proper non-extensive
%statistics can be used to study the magnetocaloric effect.
%
%
%The relation between the entropic $q$-index  concerning the generalized statistics and the entropy spectrum was
%derived using the first principles of statistics, like partition function and the definition of entropy through the specific heat construction.
%
%We have obtained an analytical solution for the entropy, which is a new result in the magnetocaloric literature.  Besides, we have calculated specific curves to analyze the behavior of our solution for several $q$'s.
%
%Since it is proven that the thermostatistical formalism can be used in self-organized critical states of large driven
%dynamical systems \cite{ts}, our proposed analytical solution for the entropy is therefore expected to hold also
%for higher-dimensional dissipative systems, to provide a close relationship between the non-extensive statistics formalism
%and more complex systems. As another perspective, we can ask if analogous considerations might exist
%for Hamiltonian magnetocaloric systems with long-range interactions.

%\newpage

\section*{Acknowledgments}

%\noindent E.M.C.A.  thanks CNPq , Brazilian scientific support federal agency, for partial financial support.

\noindent E.M.C.A.  thanks CNPq (Conselho Nacional de Desenvolvimento Cient\' ifico e Tecnol\'ogico), Brazilian scientific support federal agency, for partial financial support, Grants numbers 302155/2015-5 and 442369/2014-0 and the hospitality of Theoretical Physics Department at Federal University of Rio de Janeiro (UFRJ), where part of this work was carried out.

%\newpage

\begin{figure}[!h]
\includegraphics[width=2.7in, height=2.7in]{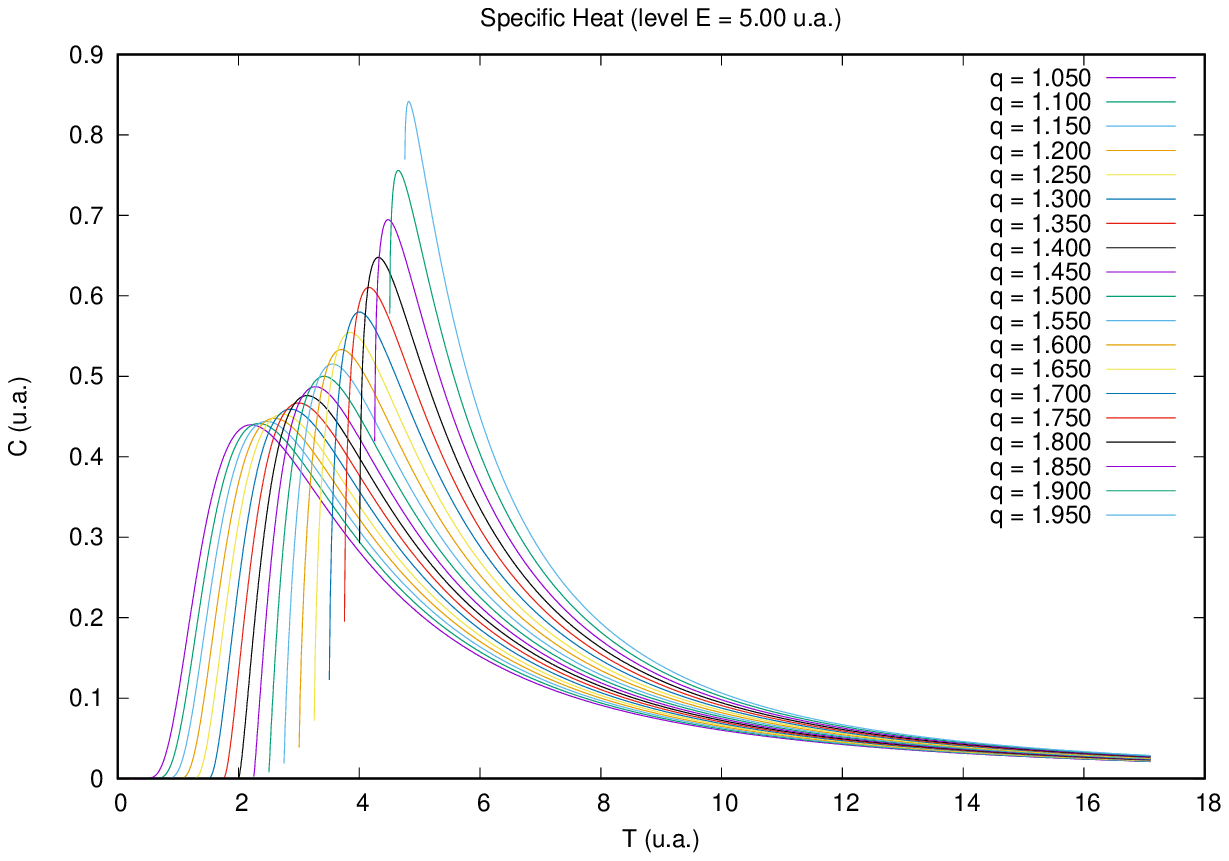}
\includegraphics[width=2.7in, height=2.7in]{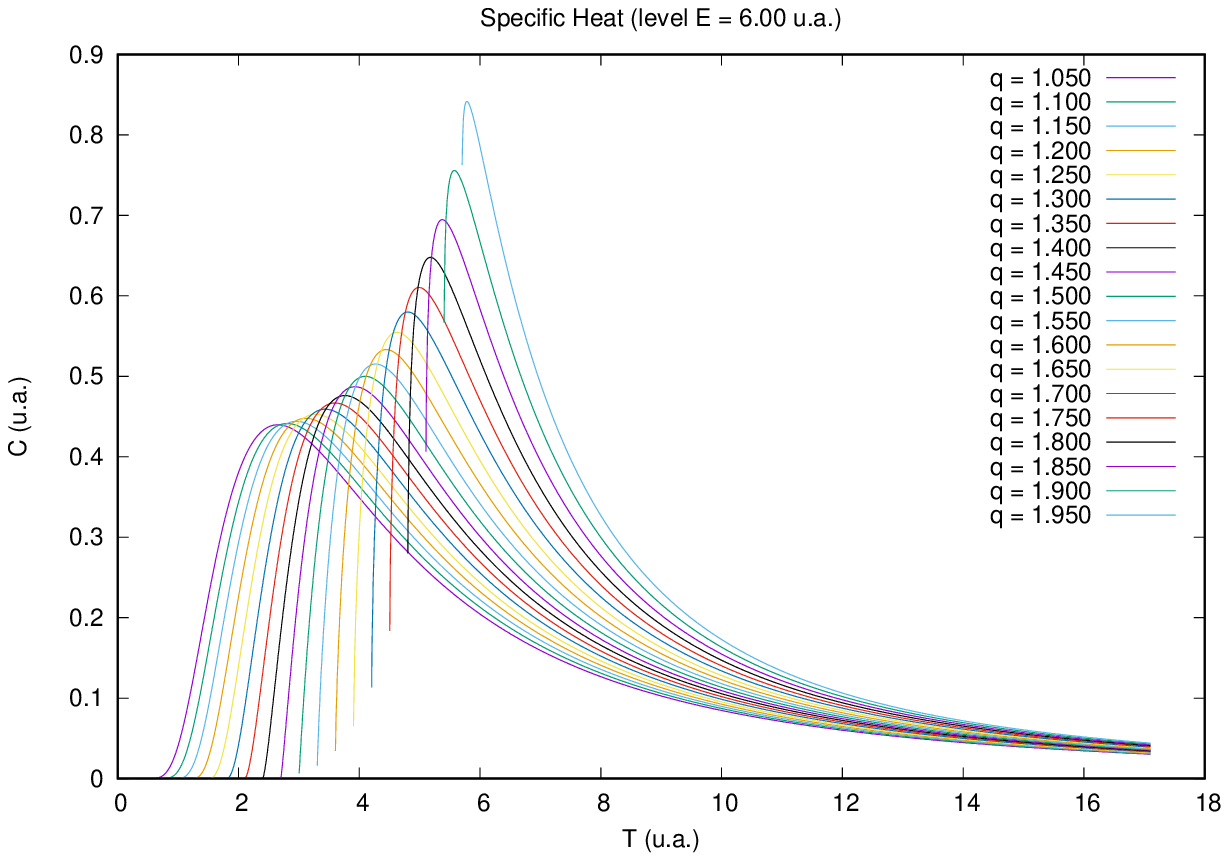}
\caption{Temperature dependence of the specific heat (in arbitrary units) for a two level quantum system for some values of the $q$ and $E=5.0$ and $E=6.0$.}
\label{specific-heat-1}
\end{figure}
\begin{figure}[!h]
\includegraphics[width=2.7in, height=2.7in]{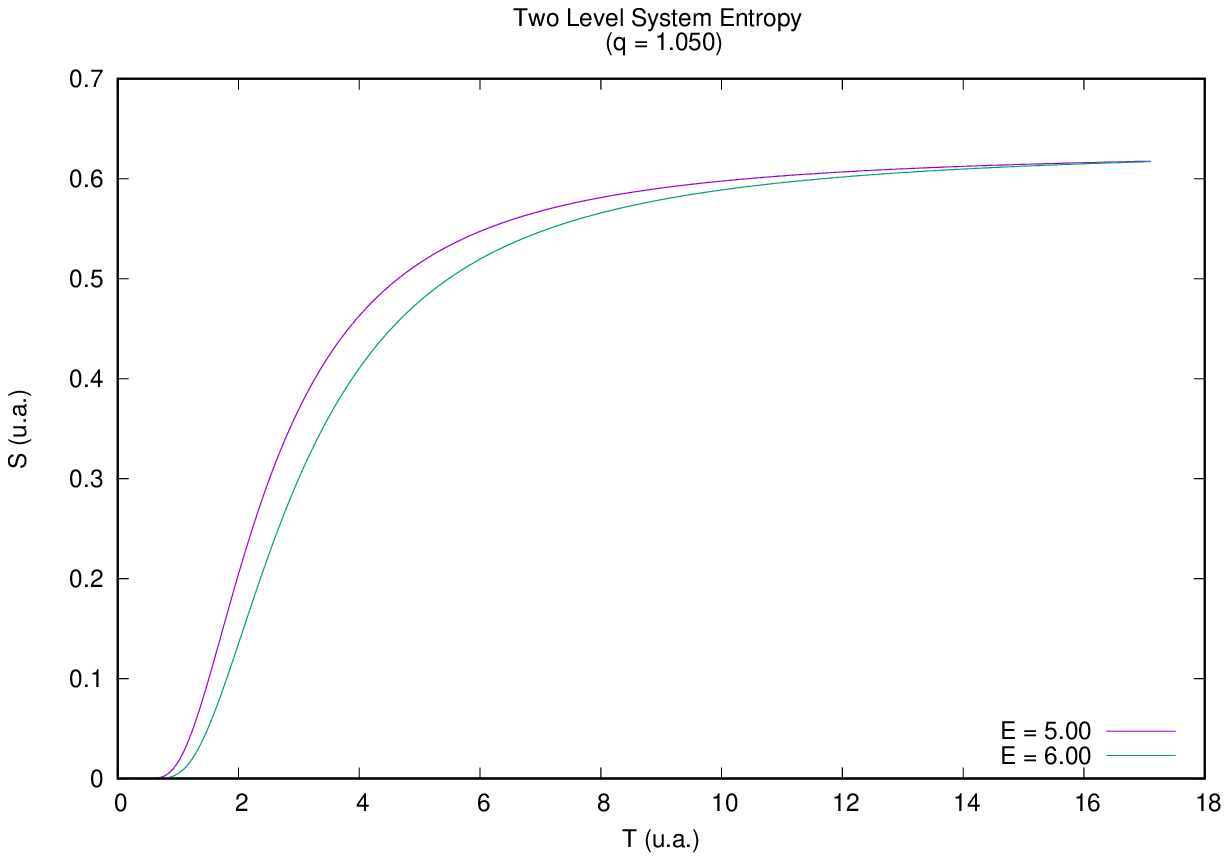}
\includegraphics[width=2.7in, height=2.7in]{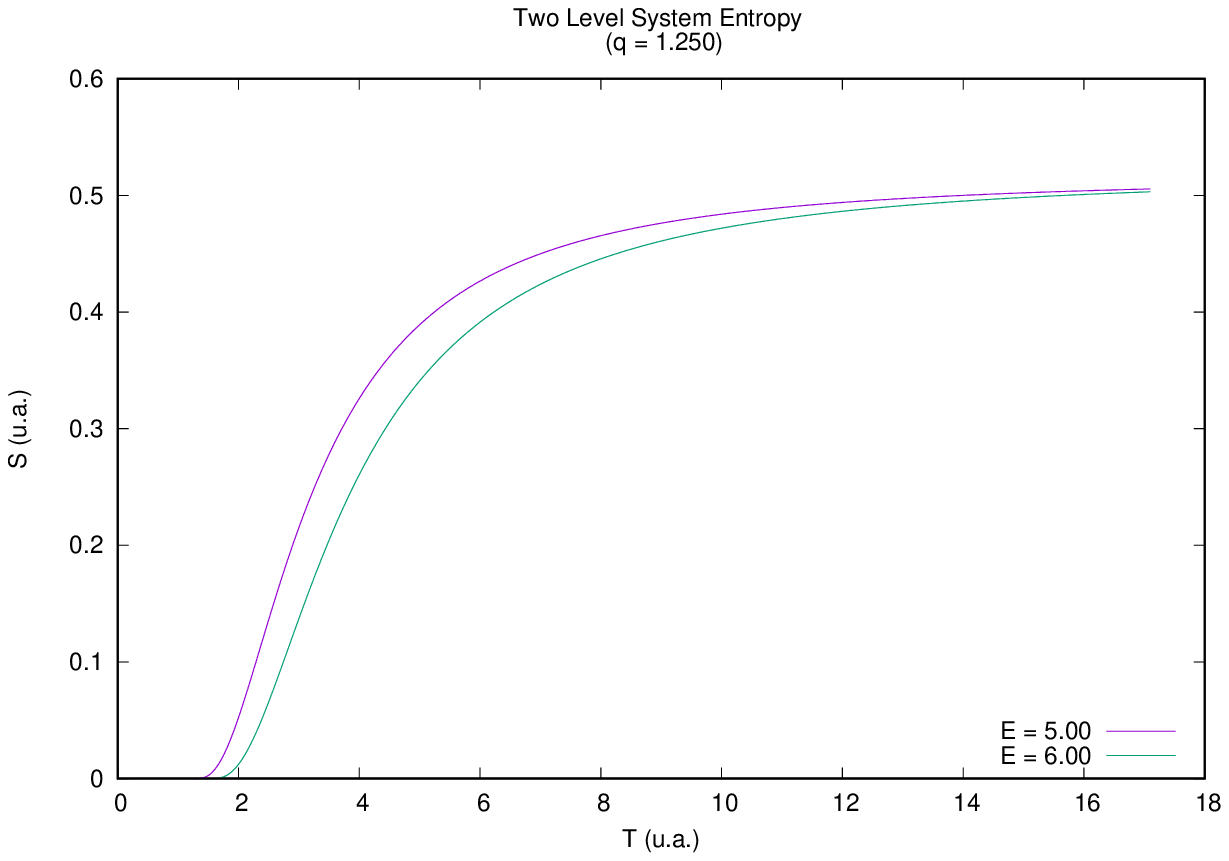}
\caption{The temperature dependence of the entropy for $E_{0}=4.0$ and $E_{1}=5.0$, when the Tsallis' parameter is $q=1.05$ and for $q=1.25$.}
\label{entropia.q01.plt}
\end{figure}
\begin{figure}[!h]
\includegraphics[width=2.7in, height=2.7in]{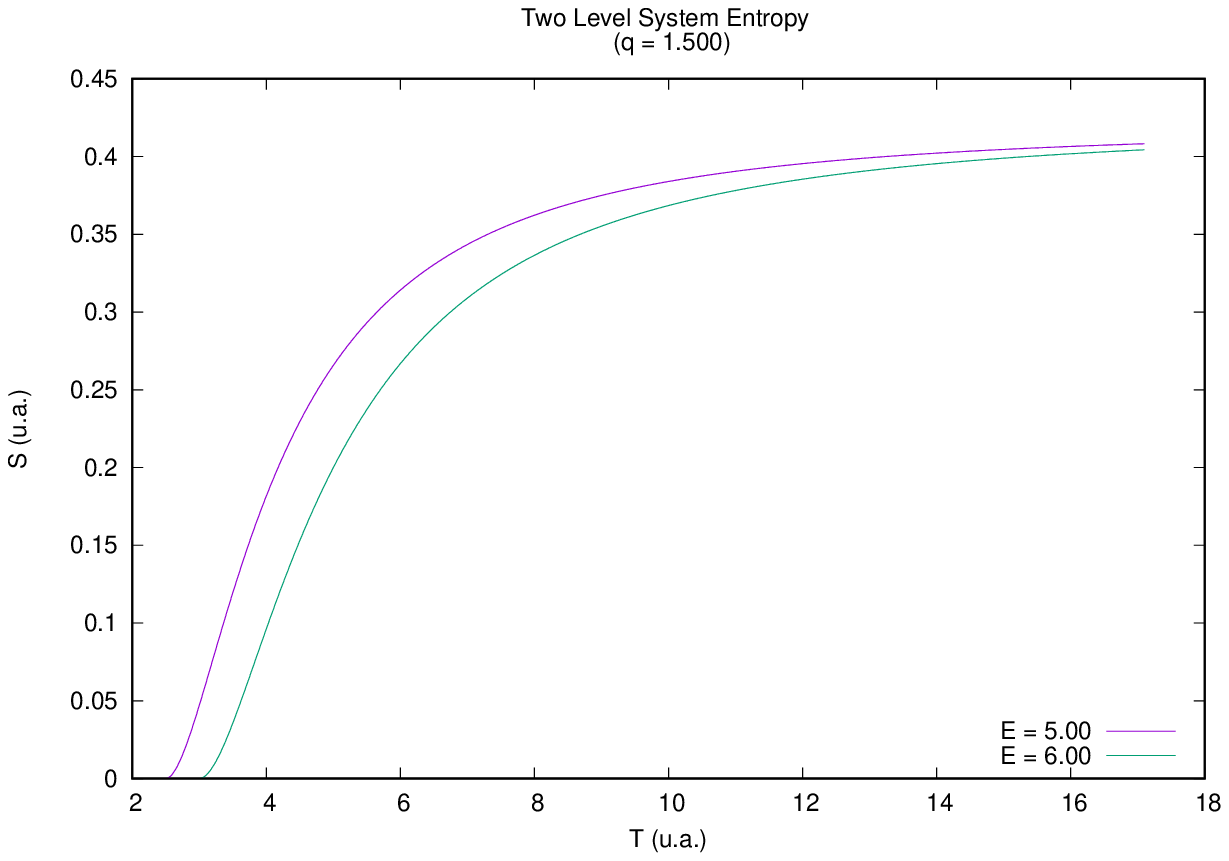}
\includegraphics[width=2.7in, height=2.7in]{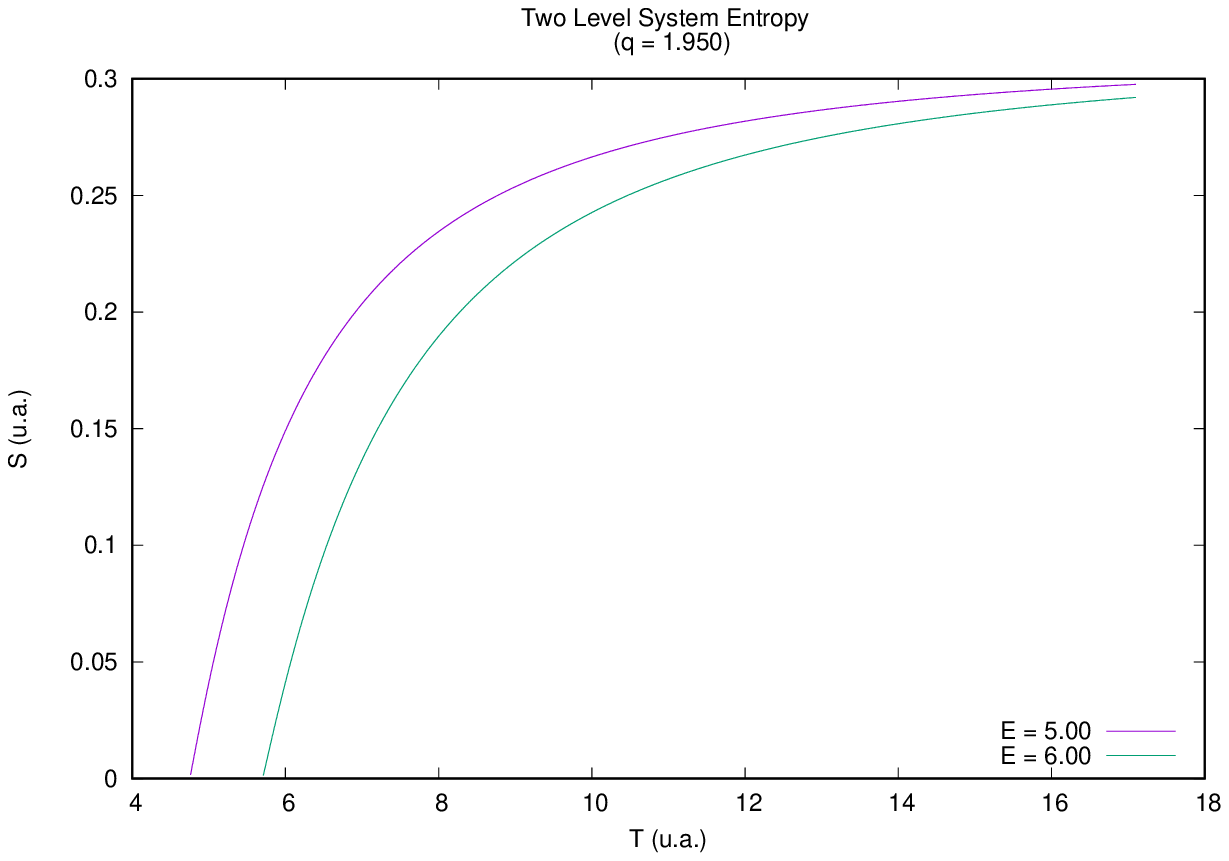}
\caption{The temperature dependence of the entropy for $E_{0}=4.0$ and $E_{1}=5.0$, when the Tsallis' parameter is $q=1.50$ and for $q=1.95$.}
\label{entropia.q10.plt}
\end{figure}

%\end{widetext}
%
%\begin{figure*}[!htbp]
%\begin{figure}[!h]
%\includegraphics[width=2.7in, height=2.7in]{DeltaQ.eps}
%\includegraphics[width=2.41in,height=2.32in]{entropia.q01.plt.eps}
%\caption{}
%\label{deltaQ}
%\end{figure}
%\end{figure*}

%

%
\begin{figure}[!h]
\includegraphics[width=2.7in, height=2.7in]{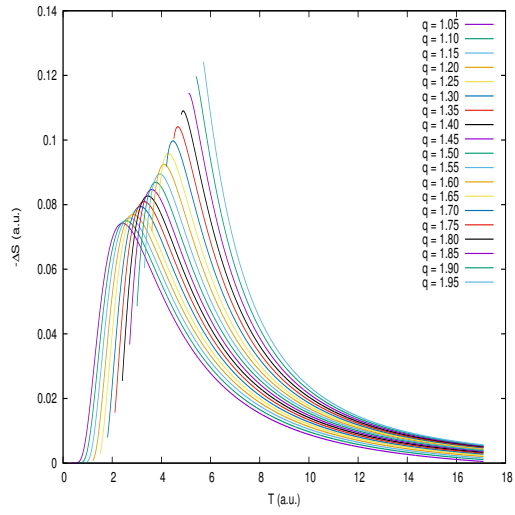}
\caption{Temperature dependence of the magnetocaloric potential $\Delta S$ for several values of the parameter $q$. Here we use the energy gap $\delta=0.5$.}
\label{deltaS}
\end{figure}

\begin{figure}[!h]
\includegraphics[width=2.7in, height=2.7in]{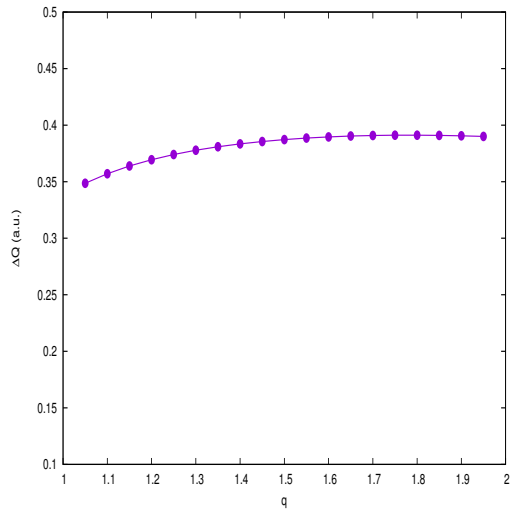}
\caption{The $q$ dependence of the heat involved on the magnetocaloric effect (MCE). The temperature range of the MCE is from $T=0$ to $T=15$ (in arb. units).}
\label{deltaQ}
\end{figure}
\begin{figure}[!h]
\includegraphics[width=2.7in, height=2.7in]{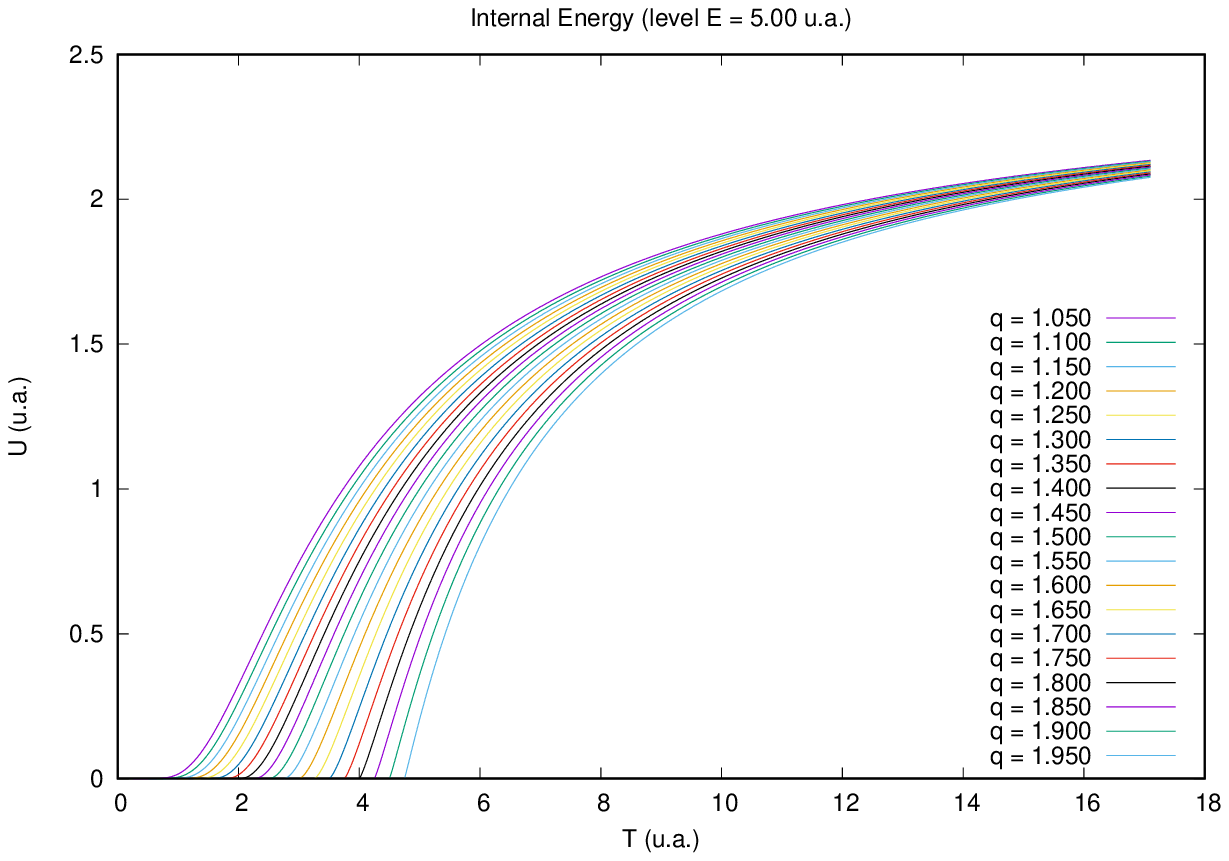}
\includegraphics[width=2.7in, height=2.7in]{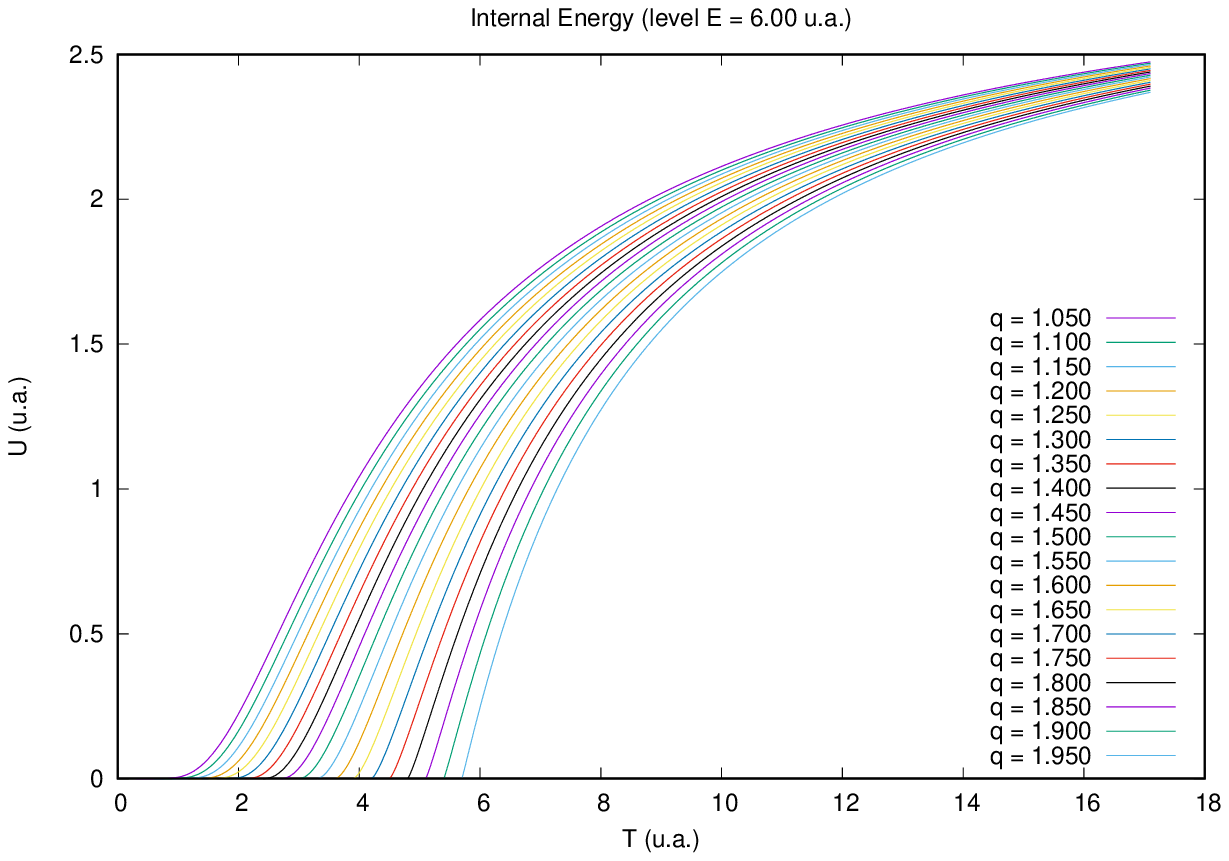}
\caption{Temperature dependence of the internal energy (in arbitrary units) for a two level quantum system for some values of the $q$ and $E=5.0$ and $E=6.0$.}
\label{energia.E1.plt}
\end{figure}
%\begin{figure}[!h]
%\includegraphics[width=2.7in, height=2.7in]{efeitoMC.eps}
%\includegraphics[width=2.41in,height=2.32in]{galfa.jpg}
%\caption{}
%\label{efeito.MC}
%\end{figure}
%
%\begin{figure}[!h]
%\includegraphics[width=2.7in, height=2.7in]{calor.E2.plt.eps}
%\includegraphics[width=2.41in,height=2.32in]{galfa.jpg}
%\caption{}
%\label{calor.E2.plt}
%\end{figure}
%
%\begin{figure}[!h]
%\includegraphics[width=2.7in, height=2.7in]{calor.E1.plt.eps}
%%\includegraphics[width=2.41in,height=2.32in]{galfa.jpg}
%\caption{}
%\label{calor.E1.plt}
%\end{figure}
%

\end{document}